\documentclass[superscriptaddress,
amsmath,amssymb,twocolumn,
aps,
pra]{revtex4-1}

\usepackage{graphicx}
\usepackage{hyperref}
\usepackage{xcolor}
\usepackage{tikz}
\usepackage{qcircuit}

\begin{document}

\title{Molecular spin qudits for quantum simulation of light-matter interactions}

\author{F. Tacchino}\affiliation{%
IBM Quantum, IBM Research -- Zurich, 8803 R\"uschlikon, Switzerland
 }

\author{A. Chiesa}
\affiliation{Universit\`a di Parma, Dipartimento di Scienze Matematiche, Fisiche e Informatiche, I-43124 Parma, Italy}\affiliation{UdR Parma, INSTM, I-43124 Parma, Italy}

\author{R. Sessoli}
\affiliation{Dipartimento di Chimica U.~Schiff, Universit\`a di Firenze, I-50019 Sesto Fiorentino, Firenze, Italy}\affiliation{UdR Firenze, INSTM, I-50019 Sesto Fiorentino, Firenze, Italy}

\author{I. Tavernelli}
\affiliation{IBM Quantum, IBM Research -- Zurich, 8803 R\"uschlikon, Switzerland
}

\author{S. Carretta}
\affiliation{Universit\`a di Parma, Dipartimento di Scienze Matematiche, Fisiche e Informatiche, I-43124 Parma, Italy}\affiliation{UdR Parma, INSTM, I-43124 Parma, Italy}\email{stefano.carretta@unipr.it}

\date{\today}

\begin{abstract}

We show that molecular spin qudits provide an ideal platform to simulate the quantum dynamics of photon fields strongly interacting with matter. The basic unit of the proposed molecular quantum simulator can be realized by a simple dimer of a spin 1/2 and a spin $S$ transition metal ion, solely controlled by microwave pulses. The spin $S$ ion is exploited to encode the photon field in a flexible architecture, which enables the digital simulation of a wide range of spin-boson models much more efficiently than by using a multi-qubit register. The effectiveness of our proposal is demonstrated by numerical simulations using realistic molecular parameters, whose prerequisites delineating  possible chemical approaches are also discussed. 

\end{abstract}

\maketitle

\section{Introduction}

In the last few years, quantum computers have emerged as a disruptive technology that promises to solve a large class of problems much more efficiently than any classical machine. The first noisy quantum processors~\cite{NISQ} are already available and enable the implementation of non-trivial algorithms targeted to specific tasks~~\cite{NatPhysIBM,vqe1,havlicek_supervised_2019,Supremacy}. In particular, thanks to their intrinsically quantum logic~\cite{RevTacchino}, they could be used already in the short term to simulate the dynamics of classically intractable quantum systems. Understanding the behavior of matter at the nano-scale is a fundamental step to design new molecules, materials and devices. However, the ``wonderful problem'' of quantum simulation ``doesn't look so easy''~\cite{Feynmann}. In fact, many examples of interest for Physics and Chemistry, such as atoms interacting with light or with thermal baths, are intrinsically difficult to be modeled on current qubit-based architectures~\cite{somma_simulating_2002}.

In this respect, Chemistry offers the change of perspective which could overcome the aforementioned difficulties. Indeed, molecular spin systems characterized by a sizeable number of accessible levels can be used to encode multi-level logical units (qudits). Each molecular qudit could replace several distinct qubits in various  algorithms~\cite{NatPhysqudit}, thus greatly simplifying manipulations of the register. Magnetic molecules are the ideal playground to implement this alternative architecture. Indeed, they are characterized by long coherence times~\cite{Freedman2014,Zadrozny,Bader,Atzori2016,Atzori_JACS,Atzori2017,Atzori2018}, which can be even enhanced by chemically designing the molecular structure~\cite{PRLWedge,Freedman_JACS} or targeting protected transitions~\cite{Hill,Freedman_Cr}.
Moreover, the spin state of these systems can be easily manipulated by microwave or radio-frequency pulses~\cite{jacsYb}, thus implementing single- and two-qubit gates in permanently coupled~\cite{Luis2011,Aguila2014,Ardavan2015} or scalable architectures, for which different ways of switching the qubit-qubit interaction were proposed ~\cite{SciRepNi,modules,Chem,VO2}. 
Recently, the idea of exploiting the additional levels typical of these systems for quantum error correction has been put forward ~\cite{jacsYb,ErCeEr,JPCLqec}. In these works, a multi-level molecule is used to encode a protected qubit within a single object, in place of the many qubits required by standard block-codes~\cite{Nielsen}.\\

Here we show how the qudit nature of magnetic molecules could simplify the practical implementation of important quantum simulation algorithms. We focus, in particular, on the simulation of light-matter interaction processes in the ultra-strong coupling regime, a problem that does not generally allow for a perturbative treatment and is therefore hard to be solved on a classical computer. This class of models, which are of crucial importance for many fundamental investigations ranging from cavity quantum electrodynamics to photochemistry~\cite{RevModPhys.91.025005,Noriultra}, have mostly been tackled so far with analog~\cite{PhysRevX.2.021007,PhysRevA.87.033814,braumuller_analog_2017} or digital-analog~\cite{lamata2918digitalanalog,lamata2020quantum} simulation strategies. The necessity to describe radiation modes (characterized in principle by infinitely many degrees of freedom) represents a major challenge for digital approaches. Standard encodings, designed for multi-qubit architectures~\cite{sawaya_resource-efficient_2020}, either employ an exponentially large Hilbert space (using a number of qubits equal to the number of simulated photons~\cite{DiPaolo}) or reduce the number of qubits at the price of non-local qubit-qubit interactions and hence complex quantum circuits~\cite{Mathis,sawaya_resource-efficient_2020}. \\

Conversely, here we reduce both the hardware overhead and the complexity of manipulations by mapping each photon mode to a single spin $S$ qudit. Thanks to the power of coordination chemistry, different qudits can be linked together and, e.g., to spin 1/2 units encoding two-level atoms ~\cite{Peng,Ni21Gd20,giant,Gd2}, in non-trivial molecular geometries. This, together with the capability of manipulating the state of the hardware by resonant and semi/resonant microwave pulses, would allow us to digitally simulate the atom-photon dynamics involving multi-mode fields and/or multiple atoms~\cite{PhysRev.93.99,DiPaolo}. \\
In particular, we show that very simple molecules consisting of dimers of transition metal ions (a spin 1/2 and a spin $S\ge 3/2$) can be used to efficiently simulate atom-photon interactions in a non-trivial range of parameters up to strong and ultra-strong coupling~\cite{RevModPhys.91.025005,Noriultra}. 
The same approach can be extended to simulate, e.g., lattice gauge  models involving many field excitations~\cite{Mathis}, by exploiting the remarkable capabilities of coordination chemistry in synthesizing multi-center molecules with very large total spin~\cite{giant}.  \\
In the following, we design the sequence of pulses allowing us (i) to determine the ground state of the simulated system using the variational quantum eigensolver algorithm (VQE)~\cite{Peruzzo2014,vqe1,vqe2,cao_review_quantum2019} and (ii) to follow the time evolution of the system prepared in an out-of-equilibrium initial state.
The remarkable performance of the proposed hardware is demonstrated by numerical simulations with parameters corresponding to existing coordination compounds~\cite{SIMqubit,Dalton2009,Abragam,Collison,Freedman_Ni,Freedman_Cr,Bader}, including the effect of decoherence and the full sequence of pulses needed to implement the algorithms.
These results make the here-proposed molecular quantum simulator very promising and pave the way to forthcoming proof-of-principle experiments.
We finally note that such qudit encoding can be easily extended to any other boson field, thus 
allowing one to simulate along the same lines many other important models, ranging from phonon vibrations~\cite{huh_boson_2015,C9SC01313J,D0SC01908A}, possibly interacting with spins~\cite{RevModPhys.59.1}, to lattice gauge theories~\cite{Mathis} and complex quantum optical setups~\cite{kottmann2020quantum}. \\

\section{Results}
\subsection{Molecular Quantum Simulator}
\begin{figure}[t]
	\centering
	\includegraphics[width=\columnwidth]{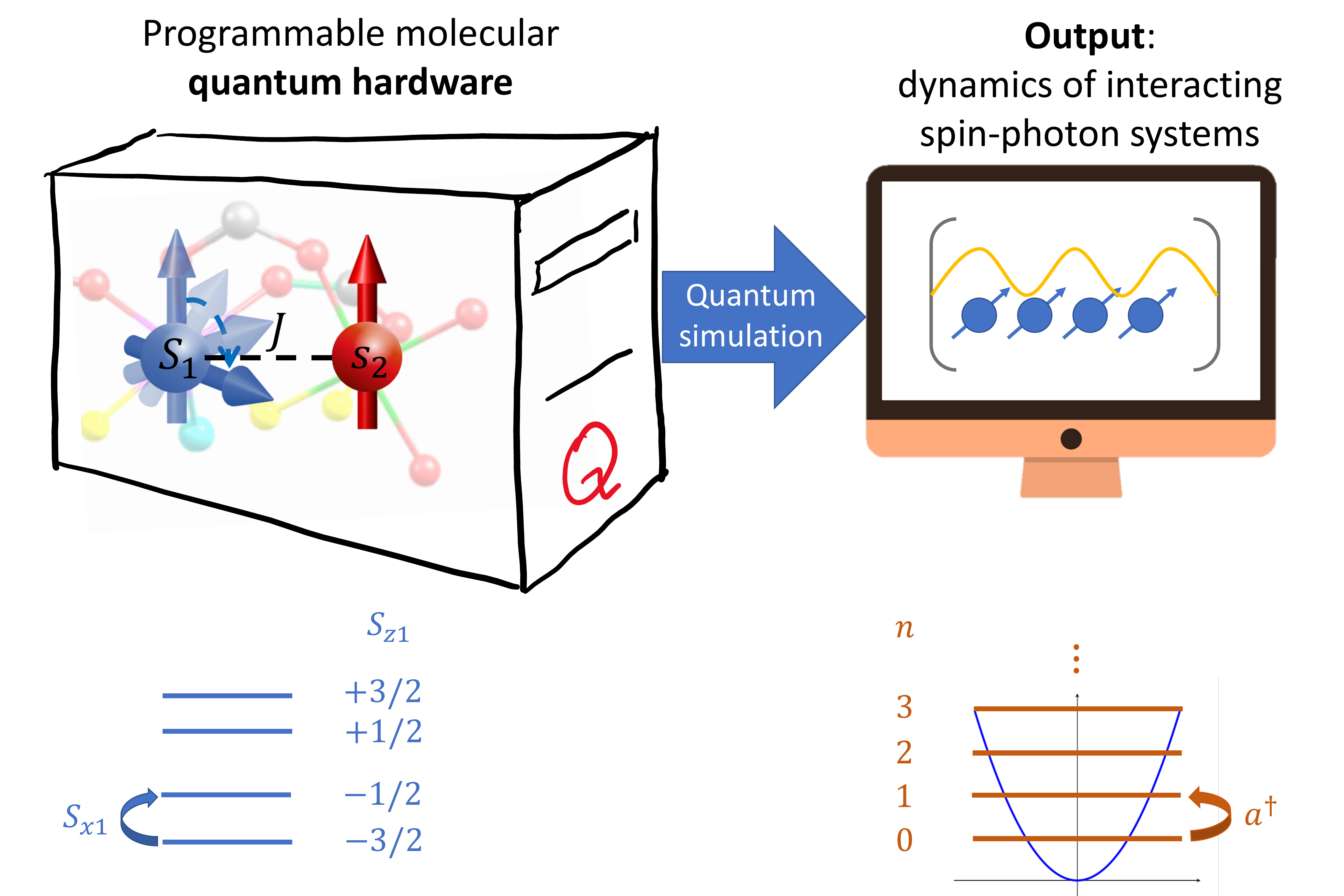} 
	\caption{Scheme of the molecular quantum simulator. Left: hardware setup, consisting of a qudit spin $S_1$ coupled by exchange interaction $J$ to a spin $s_2=1/2$. Right: target spin-photon model. In the bottom part of the figure we qualitatively sketch also the mapping between the qudit, with all the transitions made energetically distinguishable by the combined effect of Zeeman and zero-field splitting interactions, and the boson field. }
	\label{scheme}
\end{figure}

The proposed molecular hardware for quantum simulation is sketched in Fig.~\ref{scheme} (left part). 
It is a dimer consisting of a spin $S_1 \ge 3/2$ qudit that we exploit to encode the boson field, 
and an effective $s_2=1/2$, described by the following Hamiltonian:
\begin{equation}
	H_0 = g_{1z} \mu_B B S_{z1} + g_{2z} \mu_B B s_{z2} + D S_{z1}^2 + \sum_\alpha J_\alpha S_{1\alpha} s_{2\alpha}.
\label{eq:HardwareSH}
\end{equation} Here, the first two terms represent the Zeeman interaction of the two spins with an external magnetic field $B$ applied along $z$ axis and $\mu_B$ is the Bohr magneton. The third term is the zero-field splitting on the qudit (important to make all qudits transitions spectroscopically distinguishable) and the last one models an exchange or dipolar interaction between the two ions. To reduce our assumptions, we consider in the following axially anisotropic ($J_z = -2J_{x,y}$) coupling, modeling a dipole-dipole interaction between the two centers.
Different forms of the spin-spin interaction or of the single-ion anisotropy do not hinder the implementation of our scheme. The only requirement concerns the hierarchy of interactions: the transverse component of the spin-spin coupling must be much smaller than the difference between the excitation energies of the two spins $S_1$ and $s_2$. This condition guarantees that the eigenstates of Hamiltonian ~\ref{eq:HardwareSH} are practically factorized products of the states of the two spins, and can thus be labeled by $S_{z1}$ and $s_{z2}$ eigenvalues: $|\psi_{m_1m_2} \rangle \approx |m_1\rangle |m_2\rangle$. \\
These requirements are easily fulfilled in coordination compounds containing a spin 1/2 ion coupled to a spin $S_1$ transition metal ion. The latter provides the ideal qudit for the proposed architecture. As shown below, the relatively small number of levels of these qudits ($d=2S_1+1 \le 6$) is already sufficient to simulate light-matter interaction from strong to ultra-strong coupling regimes.
In addition, transition metal ion complexes with quenched orbital angular momentum ensure significantly long coherence times~\cite{PRLWedge,Bader,SIMqubit,Freedman_Cr,Freedman_Ni},  important to achieve a good simulation. 
We consider, in particular, two paradigmatic cases: Cr$^{\rm III}$ and Fe$^{\rm III}$ ions in distorted octahedral environment, yielding 
$3d^3$ and $3d^5$ electronic configurations with a single electron per orbital and thus $S=3/2$ and $5/2$, respectively~\cite{Abragam,Collison}.
Due to the practically complete quenching of the orbital angular momentum, the spectroscopic tensor $g$ is isotropic and close to the free electron value, while single ion anisotropy is typically in the $\sim 0.2-0.3$ cm$^{-1}$ range~\cite{Abragam,Collison}.
As an illustrative example, in the simulations reported below we use $D=0.24-0.30$ cm$^{-1}$ and $g=1.98$ for Cr$^{\rm III}$, as in ~\cite{PRLWedge,Freedman_Cr} and $D=-0.30$ cm$^{-1}$ and $g=2.00$ for Fe$^{\rm III}$, as reported e.g. in Ref.~\cite{SIMqubit,Zadrozny_Fe}. \footnote{By properly adjusting the static field $B$, results do not depend on the sign of $D$. } \\
These single-ion qudits can be weakly coupled through bond or through space to a spin 1/2 ion, such as Cu$^{\rm II}$ in distorted octahedral ligand cage~\cite{Abragam,Collison,Bader}, typically characterized by $g \sim 2.1-2.3$ and in some cases also by remarkable coherence times~\cite{Bader}. In the following we assume $g_{2z} = 2.3$, significantly different from $g_{1z}=1.98-2$ to ensure factorization of the system wave-function. For the dipolar interaction we assume $J_{x,y} = 0.008$ cm$^{-1}$, which corresponds to a dipolar coupling (in the point dipole approximation) between ions at a distance $\sim 6$ \AA.  A more extensive discussion on possible physical implementations is provided in Sec.~\ref{sec:phys_real}. \\
These parameters, combined with a static field of $\sim 0.3-0.5$ T ensure that $\Delta m_{1,2} = \pm 1$ transitions needed to manipulate the state of the system fall within the 20 GHz range typically explored in coplanar microwave resonators~\cite{PhysRevLett.105.140501,Rausch_2018}.

Having described in detail the molecular hardware, we now switch our attention to the target model, object of our simulation, and on how to map it onto the hardware.
The target Hamiltonian is the Rabi model~\cite{PhysRev.49.324,PhysRev.51.652,PhysRevLett.107.100401}:
\begin{equation}
\mathcal{H}_S = \omega_a \sigma_z + \Omega a^\dagger a + 2 G \sigma_x (a+a^\dagger)
\label{eq:Rabi}	
\end{equation}
Here $\sigma_z$ and $\sigma_x$ are the usual spin 1/2 operators, while $a^\dagger$ $(a)$ are bosonic creation (annihilation) operators, $[a,a^\dagger]=1$, $G$ is the atom-photon coupling, $\Omega$ ($\omega_a$) is the photon (atom) excitation energy, and we have assumed $\hbar=1$. 
Hamiltonian (Eq.~\ref{eq:Rabi}) describes the interaction between a radiation field and a two level system, such as an atom or a spin 1/2 particle. It has recently attracted a great interest in the context of quantum computing, with efforts devoted to achieve the strong coupling between superconducting qubits or spin systems and quantized photons within wave-guide resonators~\cite{PhysRevA.102.023702,gimenoAcsnano}. Behind its apparent simplicity, our target model can reveal interesting Physics and non-trivial behaviors associated to ultra or deep strong coupling regimes~\cite{Noriultra}, in which light and matter strongly mix together and exchange excitations without conserving energy~\cite{RevModPhys.91.025005}. Such a regime can also give insights into fundamental principles of lattice gauge theories~\cite{Gauge}.
We fix in the following $\omega_a=\Omega/2$ and study the model for increasing values of the $G/\Omega$ ratio, the threshold for the ultra-strong coupling regime usually being $G/\Omega\gtrsim 0.25$. \\
The molecular processor described by Hamiltonian~\ref{eq:HardwareSH} can be used to compute ground state properties and to mimic the dynamics of the target Hamiltonian~\ref{eq:Rabi}. To achieve this, we first need to encode the boson field into the spin qudit. Notice that a very good approximation can be obtained by truncating the boson field to a relatively small number of levels. %
Hence, the $d=2S_1+1$ levels of the qudit are sufficient to encode the radiation field with negligible error, by truncating the radiation field to a maximum number $n_{\rm M} = 2S_1$ of bosons. The mapping between $S_{z1}$ eigenvalues and number of bosons ($n=a^\dagger a$) is shown in the bottom part of Fig.~\ref{scheme}. In parallel, the two-state atom appearing in the Rabi Hamiltonian can be directly encoded on the hardware spin $1/2$ degrees of freedom.

Complete control of the hardware is achieved via microwave pulses resonant (or semi-resonant) with specific excitations of the spin $1/2$ or of the qudit. In particular, $\Delta m_2 = \pm 1$ transitions allow us to rotate the state of the qubit, while $\Delta m_1 = \pm 1$ pulses are used to excite the qudit. Moreover, the spin-qudit interaction enables conditioned (entangling) operations.

\begin{figure*}[ht!]
	\centering
	\includegraphics[width=\textwidth]{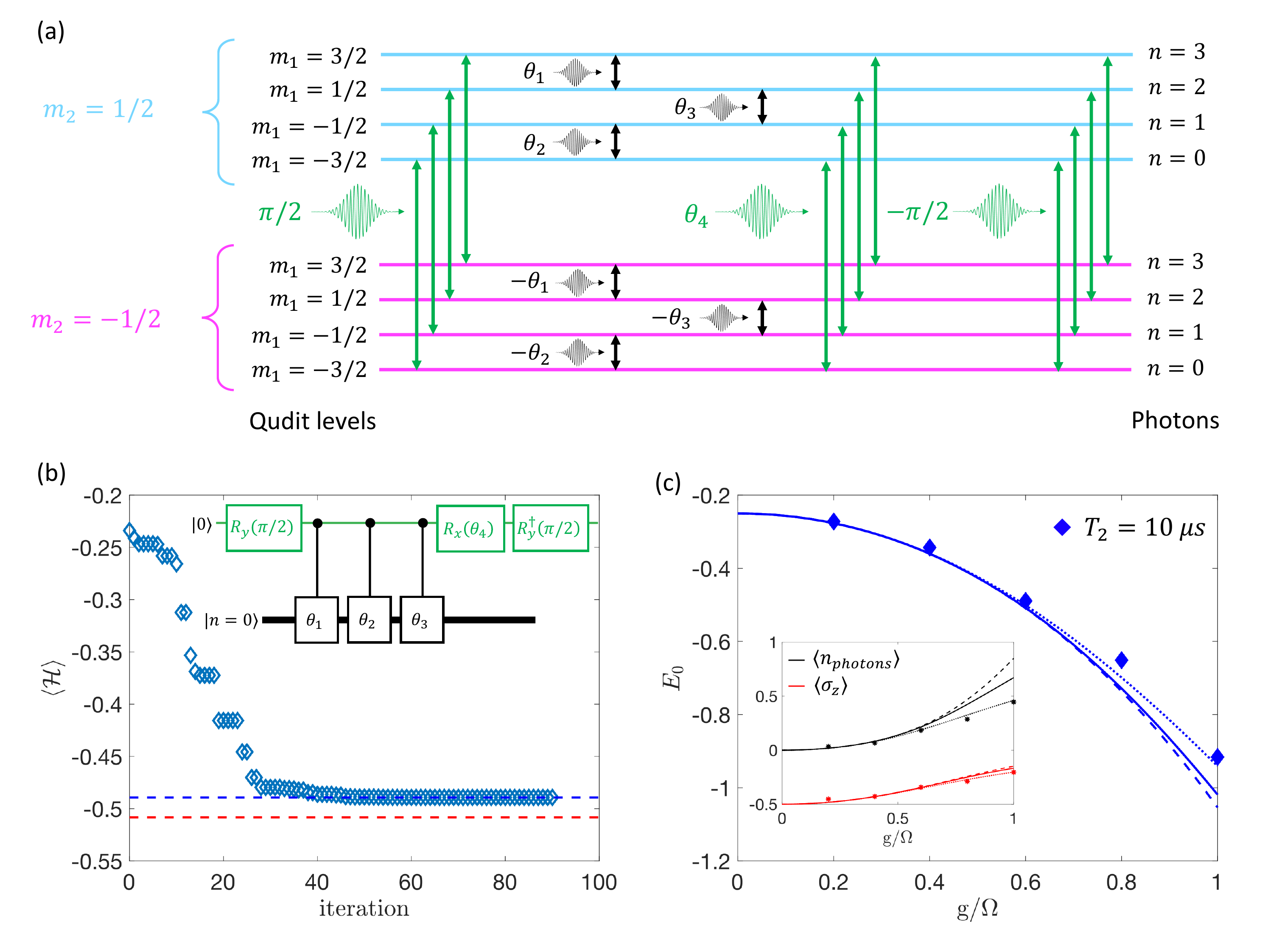} 
	\caption{Qudit-based VQE for the Rabi model. (a) Sequence of pulses for the implementation of the ground state approximation. On the left, we report the (approximate) values $m_1$ and $m_2$ for the hardware eigenstates, and on the right the corresponding photon numbers. (b) Minimization of the Hamiltonian expectation value for $G/\Omega = 0.6$. Data points converging to the value marked by the blue dashed line are obtained simulating a realistic hardware with $T_2 = 10$ $\mu$s. The red dashed line marks the optimal value achieved in the absence of errors and decoherence. The inset represents schematically the set of quantum operations which are used to approximate the ground state, including $R_{x,y}(\delta) = e^{-i \delta s_{x,y2}}$ rotations on the spin $1/2$ and conditioned $\Delta m = \pm 1$ pulses on the spin $S_1=3/2$. The variational parameters are indicated as $\theta_i$, while we assume an initial state with zero photons and a de-excited atom. (c) Ground state energy and (inset) corresponding average number of photons and atom excitations. Dashed lines represent ideal values with no approximations, solid lines are the exact results obtained by Hamiltonian diagonalization after truncating the photonic Hilbert space to $d=4$ levels, dotted lines are numerical VQE results with no errors or dechoherence and diamonds are simulations of the real device. Each expectation value $E(\vec{\theta})$ can be measured by inducing $\Delta m_{1,2} = \pm 1$ transitions with appropriate microwave pulses. Hardware parameters are $g_1 = 1.98$, $g_2 = 2.3$, $D=0.24$~cm$^{-1}$, $J_{x,y} = 0.008$~cm$^{-1}$, $B = 0.4$~T.}
	\label{fig:vqe}
\end{figure*}

\subsection{Variational quantum eigensolver}

The starting point to derive many important properties of the examined systems is the determination of its ground state wave-function.
This task can be achieved using the variational quantum eigensolver (VQE) approach~\cite{Peruzzo2014,vqe1,vqe2,cao_review_quantum2019}. This is a hybrid quantum-classical algorithm, particularly resilient to noise and therefore well suited for near term quantum processors.
It exploits the fact that the energy expectation value is minimum for the ground state of the system. The quantum hardware is used to generate an approximation of the ground state (also known as trial wavefunction or variational ansatz) for the target model, which depends on a set of free parameters $\theta_i$, and to evaluate the energy expectation value. Minimization of the evaluated energy by a classical subroutine allows us to explore the parameter space until convergence to the system ground state. It is worth noticing that this method is typically much less demanding, compared to the digital simulation of real time evolution, in terms of the complexity and length of the required sequences of quantum operations to be implemented.

Here we demonstrate an implementation of the VQE on the proposed qudit architecture applied to the target Rabi Hamiltonian, Eq.~\eqref{eq:Rabi}. We construct the trial wavefunction by designing some basic quantum operations achieved in practice via external control microwave pulses. In particular, as shown in Fig.~\ref{fig:vqe}a, we assume a $(S_1,s_2) = (3/2,1/2)$ hardware platform and we combine 
pulses resonant with transitions of $s_2$ (green arrows), implementing 
rotations of the qubit, with $\Delta m_1 = \pm 1$ pulses on the $S_1=3/2$ spin (black). To introduce entanglement in the approximate ground state, the operations on the $S_1=3/2$ spin are actually conditioned by the state of the spin $1/2$, i.e.~we rotate each pair of qudit levels by $\pm\theta_i$ depending on the sign of $m_2$ (see Fig.~\ref{fig:vqe}a). %
In total, the ansatz contains only 4 free parameters, and can be implemented with a sequence of microwave pulses that can be as fast as $\simeq 100-200$ ns. We also mention that such variational structure, which can be natively realized on our proposed qudit architecture, is closely related to the so-called polaron ansatz, which was recently implemented on superconducting quantum hardware~\cite{DiPaolo} through non-trivial decompositions into elementary qubit operations. 

In this demonstration, we combine a classical optimization routine, namely the Nelder-Mead simplex algorithm~\cite{simplexNM}, with numerical simulations of the unitary transformations corresponding to every choice of the variational parameters. In fact, each sequence of microwave pulses can be seen as the series of quantum operations reported in the inset of Fig.~\ref{fig:vqe}b. Here the black thick (green narrow) line represents the qudit (qubit). Conditioned qubit-qudit operations are depicted with black boxes, while single qubit rotations are shown in green, in direct correspondence with Fig.~\ref{fig:vqe}a.

Simulations are performed according to a realistic hardware setup, including all the required external control pulses and molecular parameters discussed above. 
The effect of a finite spin coherence time $T_2$ is included by simulating the dynamics of the hardware density matrix $\rho$ according to the Lindblad master equation~\cite{PhysRevA.87.022337}
\begin{eqnarray} \nonumber
    \frac{d\rho}{dt} &=& -i[H_0 + H_1(t),\rho] \\ \nonumber
                     &+& \frac{1}{T_2}\big( 2 S_{z1} \rho S_{z1} -S_{z1}^2\rho - \rho S_{z1}^2 \big) \\
                     &+& \frac{1}{T_2}\big( 2 s_{z2} \rho s_{z2} - \rho/2  \big)
    \label{lindblad}
\end{eqnarray}
where time dependent $H_1$ term in the Hamiltonian indicates the presence of external oscillating control fields. For simplicity, we assume the same value for the $T_2$ for both spins in the hardware. 

\begin{figure*}[ht!]
	\centering
	\includegraphics[width=0.95\textwidth]{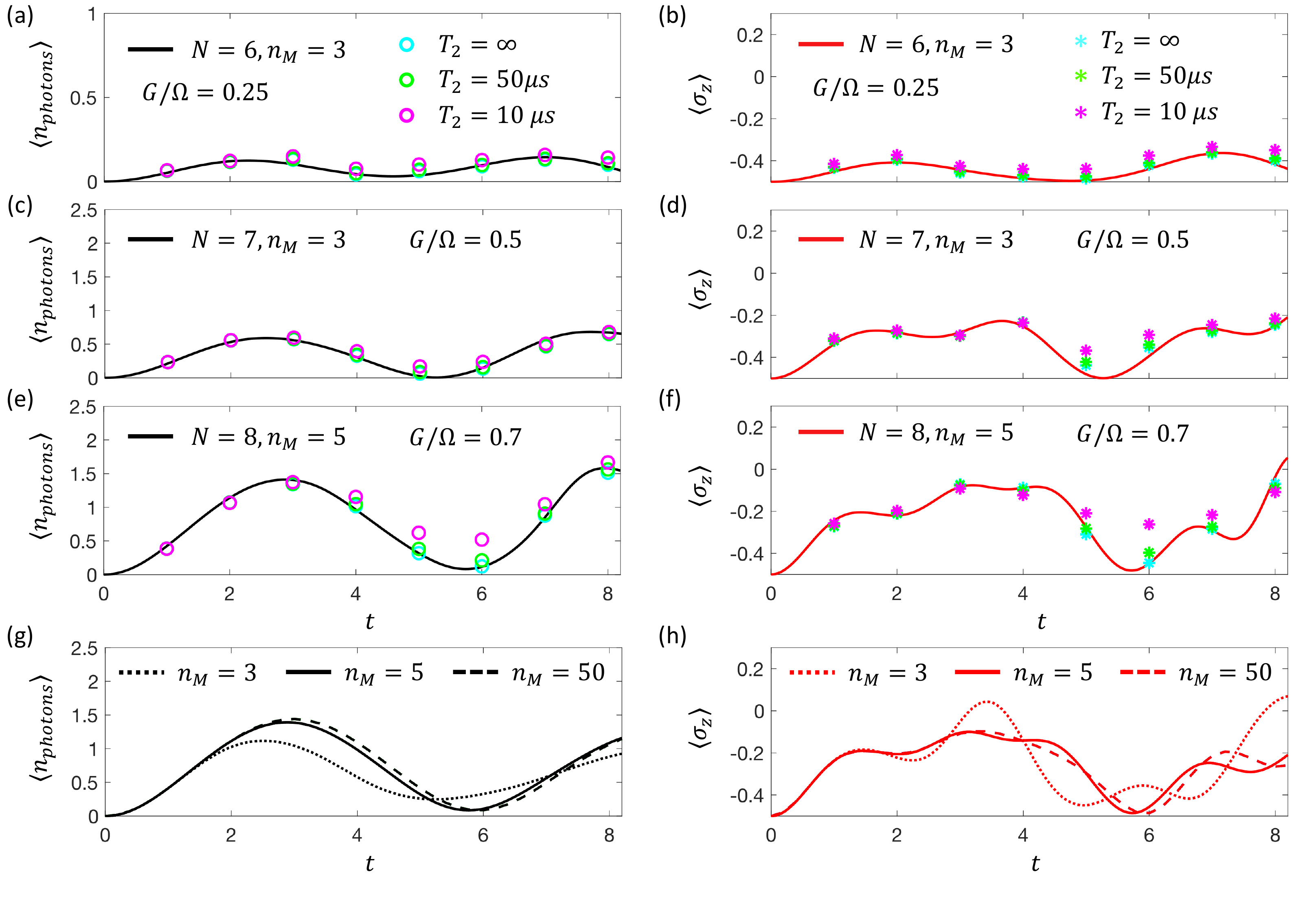} 
	\caption{Digital quantum simulation of the Rabi model on a qudit architecture. In panels (a)-(f) the solid lines are numerical results taking into account the same digital approximation and space truncation errors to which the hardware sequences are subject, but without decoherence or imperfections in the quantum operations. The data points represent hardware simulations from which we obtain the average number of photons and atom population as $\langle n_{photons}\rangle = \langle S_{z1}\rangle + S_1$ and $\langle \sigma_{z} \rangle = \langle s_{z2}\rangle$. (a-b) $G/\Omega = 0.25$ simulated with $S_1 = 3/2$, $s_2=1/2$, $g_1 = 1.98$, $g_2 = 2.3$, $D=0.24$~cm$^{-1}$, $B = 0.4$~T. (c-d) $G/\Omega = 0.5$ simulated with $S_1 = 3/2$, $s_2=1$, $g_1 = 1.98$, $g_2 = 2.18$, $D=-D^\prime=0.24$~cm$^{-1}$, $B = 0.2$~T. (e-f) $G/\Omega = 0.7$ simulated with $S_1 = 5/2$, $s_2=1$, $g_1 = 2$, $g_2 = 2.18$, $D\simeq-0.30$~cm$^{-1}$, $D^\prime=-0.24$~cm$^{-1}$, $B = 0.08$~T. (g-h) Exact time evolution ($N=\infty$) for $G/\Omega = 0.7$, showing the effect of the truncation to a maximum of $n_M = 2S_1$ photons. In all hardware simulations, we assume $J_{x,y} = 0.008$~cm$^{-1}$.}
	\label{dqs}
\end{figure*}

In Fig.~\ref{fig:vqe}c, we report results of the VQE algorithm simulated by assuming a realistic value of the spin coherence time (10 $\mu$s, symbols), compared with exact values (lines) for both the ground state energy and some ground state properties of interest. 
Notice that, over a wide range of $G/\Omega$ values, the proposed ansatz achieves very good approximations of the exact ground state. The limiting factor is essentially the expressibility of the trial wavefunction, i.e.\ the fact that by using the set of operations reported in Fig.~\ref{fig:vqe}a we may not achieve the exact form of the true ground state. This limitation can in principle be overcome by repeating the same basic parametrized structure more than once. %
It is worth noting that a finite coherence time, similarly to small imperfections in the practical realization of quantum gates, only minimally affects the final results. In fact, consistently with the underlying variational principle, noisy ground state energy estimates sometimes converge to values slightly larger than the exact ones.

\subsection{Digital quantum simulation of strong light-matter interaction}

After investigating ground state properties, we now move on to show how the proposed molecular qudit-based processor can be used to simulate the dynamics of the Rabi model. The digital quantum simulation of the target Hamiltonian $\mathcal{H}_S$ requires to implement the transformation: 
\begin{equation}
|\psi(0)\rangle \longrightarrow  |\psi(t)\rangle = e^{-i\mathcal{H}_S t} |\psi(0) \rangle.
\end{equation}
This can be approximated to the product of %
unitary terms $e^{-i\mathcal{H}_St} \approx (e^{-i\omega_a \sigma_z t/N}e^{-i\Omega_a a^\dagger a t/N}e^{-i 2G \sigma_x (a+a^\dagger) t/N})^N$ by dividing the transformation into small time steps $t/N$, according to the Suzuki-Trotter decomposition. Each unitary  step is then implemented by a sequence of micro-wave pulses. %
For instance, the effect of the diagonal operator $a^\dagger a$ is obtained by pulses semi-resonant with $\Delta m_1 = \pm 1$ transitions ~\cite{SciRep15}, while the term $\sigma_x(a+a^\dagger)$ is simulated by resonant $\Delta m_1 = \pm 1$ transitions conditioned by the state of the qubit and essentially correspond to the similar ones employed in the VQE above.

In Fig.~\ref{dqs} we show the digital quantum simulation of the Rabi model, Eq.~\eqref{eq:Rabi}, realized with the spin qudit encoding described above and for increasingly challenging choices of the $G/\Omega$ ratio. Large $G/\Omega$ values introduce peculiar features in the dynamics of the target system: the rotating wave approximation fails and the total number of excitations is not conserved. This non-trivial behavior emerges in our simulations below, where we report the time evolution of the average number of photons $\langle n_{photons} \rangle$ in the radiation mode and of the atom population $\langle \sigma_z\rangle$, assuming an initial vacuum state with zero photons and the atom in its ground state. This vacuum state (with no excitations) would not be subject to any evolution for small $G/\Omega$ ratios. Hence, oscillations in $\langle n_{photons} \rangle$ and $\langle \sigma_z \rangle$ are a direct signature of the ultra-strong coupling regime.
In all panels, we compare the reference curves, computed via exact matrix exponentiation, with numerical simulations of a realistic hardware obtained again by integrating Eq.~\eqref{lindblad}. A quantitative assessment of the overall quality of the results can be obtained by computing the fidelity $\mathcal{F} = \sqrt{\langle\psi_{id} |\rho| \psi_{id} \rangle}$ between the hardware output $\rho$ and the ideal result $\psi_{id}$ of a digital quantum simulation algorithm realized with the same number of Suzuki-Trotter time steps and the same size of the bosonic Hilbert space. The latter can be obtained by with standard matrix algebra. 

In the first example, Fig.~\ref{dqs}a-b, we show the results of the quantum simulation of the target Hamiltonian $\mathcal{H}_S$ with $G/\Omega = 0.25$ realized with $N=4$ (for $t\leq 5$) and $N = 6$ (for $t>5$) Suzuki-Trotter steps. Here, the hardware setup is composed of a spin $S_1 = 3/2$, encoding a $d=4$ photonic space, and a spin $s_2=1/2$ representing the atomic degrees of freedom. The longest pulse sequence requires $1.7$ $\mu$s, resulting in large average fidelities: $\mathcal{F} \simeq 0.984$ for $T_2 = 50$ $\mu$s and $\mathcal{F} \simeq 0.951$ for $T_2 = 10$ $\mu$s.

Increasing values of the target $G/\Omega$ ratio, Fig~\ref{dqs}c-f, yield larger oscillations in the average number of photons and atom populations. To capture these features we need, on the one hand to increase the number of digital steps ($N$), on the other hand to enlarge the bosonic space ($n_M$). This last step is fundamental to correctly capture the system dynamics at significant $G/\Omega$, as clearly shown in panels (g-h), where we compare the time evolution obtained by truncating the number of photons to 3 or 5, for $G/\Omega=0.7$. Indeed, by slightly increasing $n_M$, we practically obtain the exact dynamics (continuous line). Given $n_M=2S_1$, on the synthetic side, this simply translates in changing the qudit spin from 3/2 to 5/2. \\
Conversely, increasing $N$ (and hence the length of our manipulations) requires larger $T_2$ or faster pulses, e.g.~by engineering the molecular spectrum to better resolve all transitions. In this respect, the large degree of chemical flexibility represents a valuable resource. In particular, it is helpful to replace the $s_2 = 1/2$ with a spin $s_2 = 1$ system. A promising candidate ion is for example Ni$^{\rm II}$, for which coherence times in the regime of microseconds were reported~\cite{Freedman_Ni}. While only two consecutive levels, e.g.\ $m_2 = 0,1$, are used for the actual encoding of the target model, the presence of an additional zero-field splitting term $D^\prime s_{z2}^2$ in the hardware Hamiltonian greatly improves the frequency resolution of the relevant transitions, thus allowing for larger operation fidelities with reasonably fast control pulses. For Ni$^{\rm II}$, $D^\prime$ can be in the $0.1-1$ cm$^{-1}$ range (in octahedral ligand field)~\cite{Abragam,Collison}. In Fig.~\ref{dqs}c-d we report a digital simulation for $G/\Omega = 0.5$, obtained with $N = 7$ on a $(S_1,s_2) = (3/2,1)$ model hardware. Here, the pulse sequences last approximately $0.9$ $\mu$s on average, resulting in average fidelities $\mathcal{F} \geq 0.92$ also for $T_2 = 10$ $\mu$s. Finally, we achieve in Fig.~\ref{dqs}e-f a digital simulation well above the ultra-strong coupling threshold ($G/\Omega = 0.7$, $N=8$) with a model hardware $(S_1,s_2) = (5/2,1)$ (i.e.\ with a bosonic space truncated at $d =6$). More demanding pulse sequences are required in this case, with an average duration of $\sim 1.6$ $\mu$s and average fidelity around $\mathcal{F} \simeq 0.84$ for the shortest $T_2$.\\

\section{Possible physical implementations}
\label{sec:phys_real}

Let us now explore potential realizations of molecular qudits displaying a set of properties consistent with the ones employed in our calculations. In many cases, we refer to chemical building blocks already discussed or characterized in the literature.

To identify a suitable molecular platform, we need to combine requirements on the different units discussed in the previous sections. As already illustrated, a prototypical hardware could consist of a dimer of transition metal ions, respectively with spins $S_1 \ge 3/2$ and $s_2 \ge 1/2$. In order to ensure factorization of the two-ion wave-function, the two ions should be weakly interacting through space or through bond and characterized by $g$ factors significantly different along a given direction. Single-ion anisotropy on both $S_1$ and $s_2$ (if the latter is $\ge 1$) could for example help to better resolve different transitions. \\ 
Such single constraints do not appear so stringent. For instance, Cr$^{\rm III}$ and Cu$^{\rm II}$ have sufficiently different $g$ values $g_\text{Cr}=1.98$, $g_\text{Cu}=2.10-2.3$ to allow factorization of the wavefunction. At the same time, the individual spin’s resonance frequencies are both accessible in the same resonator. $D$ values in the order of the tenth of cm$^{-1}$ characterize ions that have half filled valence orbitals, like Mn$^{\rm II}$, Fe$^{\rm III}$, or Gd$^{\rm III}$, as well as half-filled $\text{t}_{2g}$ orbitals in octahedral ligand field, such as Cr$^{\rm III}$. It must be said that in this case the rhombicity and principal directions of the magnetic anisotropy are difficult to predict and control synthetically, but they are not crucial for the feasibility of our scheme. \\
More demanding is the control of the interaction between the spin qubit and the qudit. Dipolar interactions can be easily computed and controlled. The required strength, ca $0.01$ cm$^{-1}$, is associated to a distance of about $6$ \AA. Such a relatively short distance inside a molecular architecture is compatible with compact linkers like oxalate, cyanide, azide etc. These bridging ligands are very efficient in transmitting also exchange interactions, and thus unsuitable for single spin addressing. The optimal choice falls on very weak exchange interactions that are expected to be almost ubiquitous when the two spin centers are embedded in the same molecular scaffold. However, such weak interactions (of the same order of inter-molecular ones) have been poorly characterized through standard magnetometry techniques in concentrated solids or ab initio calculations. An elucidating example of the wide range of achievable interactions is the case of condensed Cu$^{\rm II}$ porphyrin complexes. Complexes with conjugated macrocyclic ligands have been attracting increasing interest for the relatively long and robust coherence combined with semiconducting properties and convenient processability~\cite{warner_potential_2013,Atzori2016,Bader2016,Urtizberea2018}. Electron–electron double resonance has been recently used to investigate the spin-spin interactions in edge-fused coplanar Cu$^{\rm II}$ dimers and in meso–meso singly linked dimers~\cite{Wili2019}. In the latter, the Cu-porphyrin rings are mutually orthogonal and exchange interaction fully suppressed, significantly smaller than the dipolar interaction, estimated to be $0.0028$ cm$^{-1}$. On the contrary, the planarity imposed by the triple link between the two units boosts the antiferromagnetic exchange interaction to $J=-2.64$ cm$^{-1}$. The difficulty to predict the actual spin-spin interaction (based on ab-initio calculations or simple geometrical considerations) does not hinder the implementation of our scheme. Indeed, once the complex has been characterized, it is possible to tune the external field in order to ensure factorization of the wave-function. This could require to adapt the experimental setup to work at larger frequencies than commercial resonators, as demonstrated for instance in Ref.~\cite{Rausch_2018}, where superconducting coplanar resonators operating up to $50$ GHz were reported. These superconducting resonators could also employ high-$T_c$ superconductors to support large magnetic fields~\cite{Ghirri2015}. \\   
The choice of the linkers between the two magnetic ions should also fulfil other constraints. In particular, we need to control the decoherence of the system. A coherence time $T_2$ of 10 $\mu$s at low temperatures is often observed for $S=1/2$ transition metal ions, especially if the first coordination sphere is nuclear spin free, e.g.\ oxygen, sulphur, or carbon donor atoms, and if total or partial deuteration of the ligand is affordable. This requires to eliminate nitrogen from the first coordination sphere and aliphatic ${\rm CH}_n$ groups in the molecular scaffold, thus reducing the available library of molecular candidates.\\
We should finally remind that an efficient operation of the simulator requires that the qudit-qubit pairs are well isolated, still retaining a control over the molecular orientation. An isostructural diamagnetic matrix is thus mandatory. While this is usually accessible for single qubits, in the case of a two-spins architecture the co-crystallization of the para- and dia-magnetic molecules must occur without metal scrambling. This can be easily avoided using inert $d^3$/$d^6$ ions, as in the case of Cr$^{\rm III}$ and low spin Co$^{\rm III}$. Metal scrambling is however much more common for labile $d^1$/$d^9$ ions, such as Cu$^{\rm II}$, requiring the use of polydentate linkers in the design of the molecular architecture.

\section{Discussion and conclusions}

Summarizing, we have shown that magnetic molecules are very promising quantum simulators for complex physical systems, in particular for target Hamiltonians involving bosonic variables representing e.g. radiation fields. The many degrees of freedom present in this class of target Hamiltonians make their simulation with a multi-qubit register very demanding, both in terms of number of qubits and sequence of operations. In contrast, the multi-level structure typical of magnetic molecules allow us to encode a boson into a single spin qudit, thus greatly simplifying the architecture of the register and its manipulation. The latter can be achieved solely by sequences of microwave pulses, resonant with specific transitions. As an example, we have reported both ground state calculations, performed with the VQE algorithm, and the digital quantum simulation of real time evolution for the Rabi model up to the ultra-strong coupling regime. In all cases, the outcomes obtained by considering realistic hardware parameters are in very good quantitative agreement with exact predictions.\\
These results pave the way to proof-of-principle experiments demonstrating the effectiveness of our proposal. The scheme is flexible and allows one to simulate a wide range of interesting models, thanks to the chemical tunability of the proposed hardware. Indeed, although we have focused here on very simple single ions, much larger $S_1$ can be obtained by exploiting the total spin ground multiplet of multi-nuclear complexes with tailored interactions~\cite{giant,Mn19powell}. With larger $S_1$, one could for example include more photons in the simulations, thus enabling the treatment of more exotic regimes such as the deep strong coupling for light-matter interactions~\cite{PhysRevLett.105.263603} or fundamental models such as lattice gauge theories~\cite{Montangero_lattice_2016,Carmen_2020,Mathis}. The latter require a large number of boson modes and excitations for a detailed description in arbitrary dimension, thus representing a challenging task for both classical devices and near time qubit-based architectures~\cite{Mathis}.
Additionally, models involving multiple two-level atoms or boson modes~\cite{PhysRev.93.99} can be simulated by chemically engineering the structures in order to link together several qudit and/or qubits~\cite{Peng,Ni21Gd20}.

In conclusion, it is worth stressing that a synthetic effort to achieve the conditions highlighted in this work would place molecular nanomagnets among the most promising platforms for the realization of effective quantum simulators.

\section*{Acknowledgements}
We thank P.~Santini for useful discussions. This work has received funding from the European Union's Horizon 2020 research and innovation programme through FET-OPEN grant 862893 FATMOLS and QUANTERA project SUMO (co-funded by Italian Ministry of University and Research). \\

%

\end{document}